\documentclass[letter]{aa} 

\usepackage{graphicx}
\usepackage{subfigure}
\graphicspath{ {images/} }

\usepackage{txfonts}
\usepackage{booktabs}
\usepackage{amsmath}
\usepackage{epstopdf}
\usepackage{enumitem}
\usepackage{multicol}
\usepackage{multirow}
\usepackage{chngcntr}

\usepackage{natbib}
\begin{document} 

\title{Sub-arcsecond imaging of Arp\,299-A at 150 MHz with LOFAR: Evidence for a starburst-driven outflow}
\titlerunning{Evidence for a starburst-driven outflow in Arp299-A observed by LOFAR}
\authorrunning{N. Ram\'irez-Olivencia et al.}

   \author{N. Ram\'irez-Olivencia \inst{1}
          E. Varenius \inst{2},
          M. P\'erez-Torres \inst{1,3},
          A. Alberdi \inst{1},
          E. P\'erez \inst{1},
          A. Alonso-Herrero \inst{4},
          A. Deller \inst{5},
          R. Herrero-Illana \inst{6},
          J. Mold\'on \inst{2},
          L. Barcos-Mu\~noz \inst{7,8},
          I. Mart\'i-Vidal \inst{9}
          }

   \institute{$^1$ Instituto de Astrof\'isica de Andaluc\'ia (IAA-CSIC),
              Glorieta de la Astronom\'ia s/n, 18008 Granada, Spain\\
              $^2$ Jodrell Bank Centr for Astrophysics, Alan Turing Building,
              University of Manchester, Oxford Road, Manchester M13 9PL, United Kingdom \\
              $^3$ Visiting Scientist: Facultad de Ciencias, Univ. de Zaragoza, Spain\\
              $^4$ Centro de Astrobiolog\'ia (CSIC-INTA), ESAC Campus, 28692 Villanueva de Cañada, Madrid, Spain \\
              $^5$ Centre for Astrophysics \& Supercomputing
              Swinburne University of Technology, John St, Hawthorn VIC 3122 Australia \\
              $^6$ European Southern Observatory, Alonso de C\'ordova 3107, Vitacura, Santiago de Chile \\
              $^7$ Joint ALMA Observatory, Alonso de C\'ordova 3107, Vitacura, Santiago, Chile \\
              $^8$ National Radio Astronomy Observatory, 520 Edgemont Road, Charlottesville, VA 22903, USA \\
              $^9$ Department of Space, Earth and Environment, Chalmers University of Technology,     Onsala Space Observatory, 439 92 Onsala, Sweden \\
              }

   \date{Received MM/DD, YY; accepted MM/DD, YY}


\abstract{We report on the first sub-arcsecond (0.44 $\times$ 0.41 arcsec$\rm ^2$) angular resolution image at 150 MHz of the A-nucleus in the luminous infrared galaxy Arp\,299, from International Low Frequency Array (LOFAR) Telescope observations. 
The most remarkable finding is that of an
intriguing two-sided, filamentary structure emanating from the A-nucleus,
which we interpret as an outflow that extends up to at least 14
arcseconds from the A-nucleus in the N-S direction ($\rm \approx 5$ kpc deprojected size) and accounts for almost 40\% of the extended emission of the entire galaxy system. 
We also discuss HST/NICMOS [FeII] 1.64$\rm \,\mu m$ and H$_2$ 2.12$\rm \,\mu m$ images of Arp\,299-A, which show similar features to those unveiled by our 150 MHz LOFAR observations, providing strong morphological support for the outflow scenario. Finally, we discuss unpublished Na\,I D spectra that confirm the outflow nature of this structure. 
From energetic arguments, we rule out the low-luminosity active galactic nucleus in Arp\,299-A as a driver for the outflow. On the contrary, the powerful, compact starburst in the central regions of Arp\,299-A provides plenty of mechanical energy to sustain an outflow, and we conclude that the intense supernova (SN) activity 
in the nuclear region of Arp\,299-A is driving the observed outflow. 
We estimate that the starburst wind can support a mass-outflow rate in the
range (11 - 63) $\rm M_{\odot}yr^{-1}$ at speeds of up to (370 - 890) $\rm km s^{-1}$, and is relatively young, with an estimated kinematic age of $\rm (3 - 7)$ Myr.
Those results open an avenue to the use of low-frequency ($\rm 150$ MHz), sub-arcsecond imaging with LOFAR to detect outflows in the central regions of local luminous infrared galaxies. 
}


\maketitle

\section{Introduction}   
 
Many starburst galaxies show a low-frequency turnover (below 1.4 GHz) in their spectral energy distribution (SED) \citep{Condon92}, which carries relevant information on the absorption processes taking place in those galaxies. However, until the advent 
of the International Low Frequency Array Telescope (LOFAR; \citealt{LOFAR2013}), those studies lacked the necessary angular resolution and sensitivity at $\rm 150\, MHz$ to study the low-frequency morphology of nearby luminous infrared galaxies (LIRGs) in detail. Using LOFAR at this frequency we have studied two of the most interesting starburst galaxies in the nearby universe: the closest starburst galaxy M82 \citep{varenius2015}, and the closest ultra-luminous infrared galaxy (ULIRG) Arp\, 220 \citep{varenius2016}. In this letter we present new $\rm 150\, MHz$ observations of the most luminous LIRG  within 50 Mpc: Arp\,299.

The Arp\,299 galaxy is located at $D \rm \approx 45\, Mpc$ and exists as an interacting galaxy pair   (NGC\, 3690A and NGC\, 3690B, eastern and western galaxies, respectively) in an early stage of merging. With an infrared luminosity of $\rm \log (L_{IR}/L_{\odot}) = 11.88$, it is the brightest known LIRG within 50 Mpc in the northern hemisphere.  Radio interferometric observations of Arp\,299 at centimeter wavelengths obtained with arcsecond resolution show that the two main components of the galaxy system also host the brightest radio-emitting regions (e.g., \citealt{Neff}). Moreover, very long baseline interferometry has unveiled prolific supernova (SN) factories in the central regions of Arp\,299-A \citep{Perez-Torres09,Ulvestad,Bondi}. This central SN factory fits well with the observed behavior in merger systems, which have more molecular gas ($\rm \approx 3 \times 10^9\, M_{\odot}$, \citealp{Rosenberg}) than  non-interacting galaxies \citep{combes94,casasola04}. It has also been proved that this gas tends to be concentrated in the nuclear regions of the merging galaxies \citep{Braine}.
 
Here, we present the first results from our observations of Arp\,299, focusing on the A-nucleus, where we have found evidence for a nuclear outflow powered by the intense starburst activity in the central kpc region of Arp\,299-A.The outflow is detected at these radio wavelengths since it transports synchrotron-emitting cosmic rays (CRs) and almost the whole of its structure is unaffected by absorption.


\section{Observations}
\label{sec:reduction}
We present data taken with LOFAR at $\rm 150\, MHz$ for 12 hours during the night between 22 and 23 February 2016 under project code LC5\textunderscore20 (P.I. P\'erez-Torres, M.). Three simultaneous beams were used with the same 32~MHz frequency coverage, centered at 150~MHz, to observe the target Arp\,299 and the two calibrators J1127+5841 (calibrator for phase, $\rm 10'$ from Arp\,299) and J1128+5925 (calibrator for delay/rate/amplitude/bandpass, $\rm 0.8^\circ$ from Arp\,299). Every 30~min the observations switched to a single beam on the absolute flux density calibrator 3C~295 for 3~min; see Table \ref{tab:targetlist} for source positions. The data were correlated in linear polarization (XX,XY,YX,YY) and stored in the LOFAR Long Term Archive (LTA) averaged to 4~sec in time and 4 channels per sub-band (i.e., a frequency resolution of 48.75~kHz). This limits the 10\% smearing radius for international baselines to about $\rm 5'$ around the respective beam centers for 1\,000~km baselines. For a more detailed explanation of the reduction process with LOFAR, see \ref{appendix}.

In addition, for the NaI line absorption analysis, we also present INTEGRAL, a fiber-based integral field system \citep{Arribas} and observations of 
Arp\,299-A that were performed on 3 May 2014. INTEGRAL is connected to the Wide Field Fibre Optic Spectrograph (WYFFOS; \citealt{Bingham}) attached to the 4.2m WHT in La Palma. The set up was the R120, the four exposures from the ING (Isaac Newton Group) archive at the  Cambridge Astronomy Survey Unit (CASU) of the Astronomical Data Centre (ADC). The exposures were median combined and different spectra were extracted from the nucleus and from an adjacent region. 


\section{Results}

\begin{figure}[h!]\centering   \includegraphics[width=.5\textwidth]{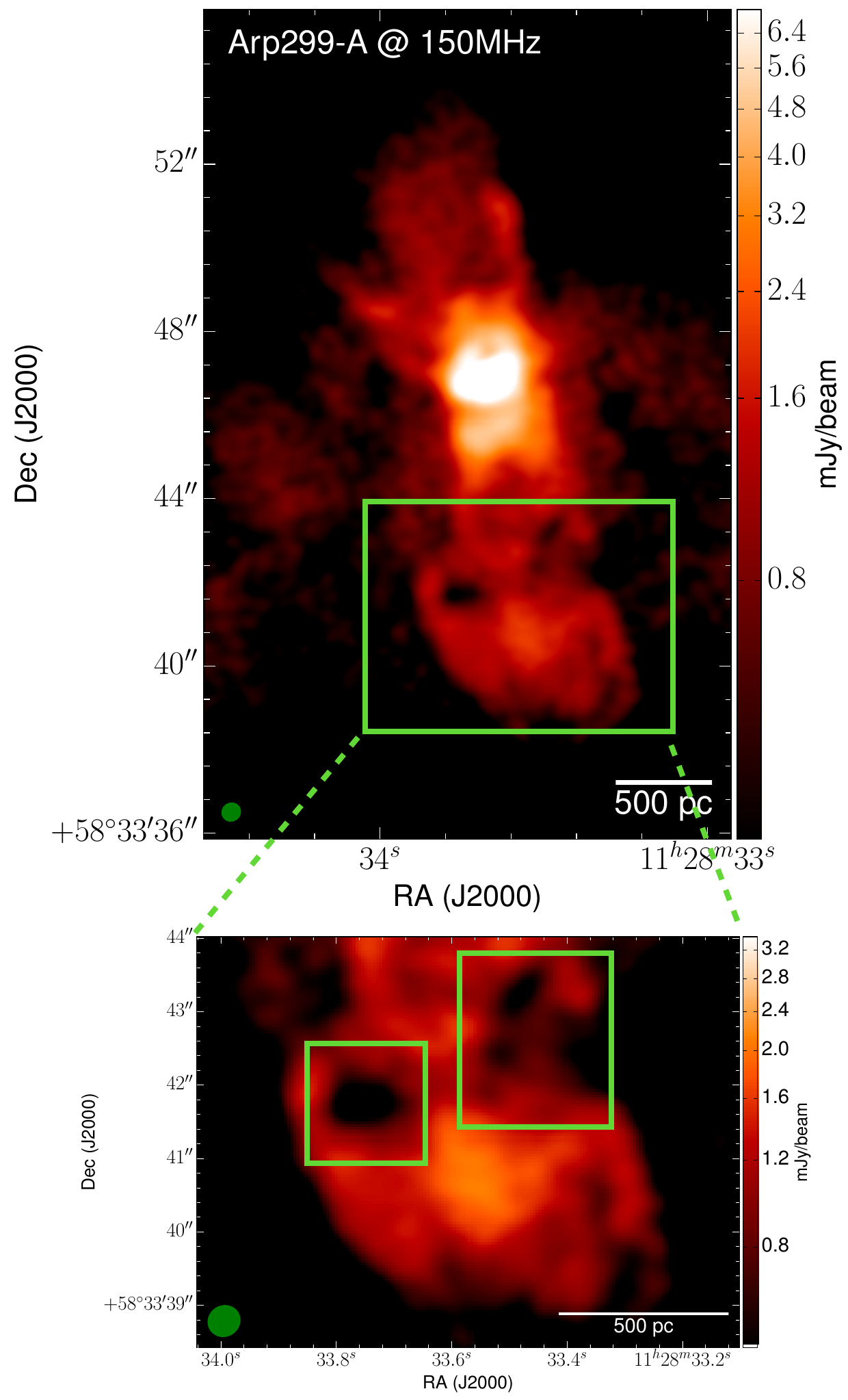}
\caption{\textit{Upper panel:} LOFAR Image of Arp\,299A at 150MHz with a beam of 0.44 $\times$ 0.41 arcsec$^2$ (Position Angle = $\rm -51.88^{\circ}$) and 
a sensitivity of $\rm rms = 90\,\mu Jy\,beam^{-1}$. \textit{Lower panel:} Zoomed image of the region of $\rm \approx 1.2 \times \,1.6$ kpc marked with a green box in 
the upper panel. The absorbed regions of the southern part of the outflow are indicated with green boxes. The sizes are approximately $\rm 330\,pc \times 380$\,pc (left) and $\rm 420\, pc \times 500$\,pc (right).} 
\label{fig:LOFAR}
\end{figure}

We show in Figure \ref{fig:LOFAR} the Arp\,299-A image ($\rm 0.44 \times 0.41\, arcsec^2$) obtained from our LOFAR $\rm 150\, MHz$ observations. The rms achieved is $\rm \approx 90\, \mu Jy\, beam^{-1}$. The peak of brightness of the image is located at R.A. = $\rm 11^h28^m33.487^s$  Dec. = $\rm +58{\circ} 33'47.044''$ (J2000.0), with a value of $\rm 21.5\, mJy\, beam^{-1}$, and corresponds to the brightest point in the entire Arp\,299 system.

The low-surface-brightness component (which we define as regions with signal-to-noise $\rm 3\sigma < S_{150\, MHz} < 50\sigma$) of the image accounts for 40\% of the extended emission of the entire galaxy, and is about four times more luminous than the A nucleus (which we define as the region with signal-to-noise $\rm > 50 \sigma$.) 
The distribution of this emission forms a biconical structure with an extension $\rm \approx 14\, arcsec$ in length and a maximum width $\rm \approx 8\, arcsec$. The northern region shows edge-brightening. However, in the southern region this behavior has not been detected, showing a more uniform emission distribution.

\section{Discussion}{\label{sec:discussion}}

\subsection{An outflowing wind unveiled with LOFAR?}

The most remarkable finding of our 150 MHz LOFAR observations is an
intriguing two-sided, wide filamentary structure emanating from the A-nucleus that extends up to at least 14 arcseconds ($\simeq 3$ kpc) from the A-nucleus in the N-S direction. The peculiar shape of this extended emission is very suggestive of an outflow (e.g., \citealt{Su,Barcos-Munoz,Mckinley}). 

The presence of obscured regions seen only in the Southern part, at a distance of $\rm \approx 1.1 kpc$ from the A nucleus (see green boxes in Figure \ref{fig:LOFAR}), suggests the presence of a nonuniform (clumpy) distribution of absorbers over the emitter. This is consistent with the existence of a spiral arm of the galactic disk seen before in, for example, Figure 9, left panel, second row \citet{Alonso-Herrero}. The northern part of the structure appears less obscured. This absorption structure is consistent with a bipolar structure emanating from the center of an inclined absorbing galactic disk (Figure \ref{fig:toy_model}). Considering that the disk may host star-forming regions with ionized gas (HII regions), these would produce free-free absorption \citep{Condon92}, being  the natural cause of the absorption at $\rm 150\, MHz$. This behavior also indicates that, although the predominant star formation activity is located at the nuclei of the interacting galaxies, ongoing star formation is also present in the spiral disk. This was also previously suggested for Arp299 by  \citet{Alonso-Herrero} in particular, and in other interacting galaxies in general (M81 of the M81 group \citealt{Kaufman,casasola07}; NGC3627 of the Leo Triplet \citealt{Smith,Paladino}). Therefore, considering that what is seen here is a unique structure crossing the nucleus, and together with the filamentary shape of the northern (nonabsorbed) region, everything suggests that this structure is an outflow.  

\subsection{Ancillary data at other wavelengths}

\textit{NICMOS/HST [FeII] and $H_2$ line emission}: In Figure \ref{fig:RGB_FeIIH2} we present the $\rm H_2$ ($\rm 2.12 \mu m$) and FeII ($1.644 \mu m$) 
emission-line images from \citet{Alonso-Herrero}. The images represent the same region of Arp\,299-A, using a logarithmic color scale to enhance the faint, extended emission. The structures present in both images also trace a similar bi-conical behavior of the extended emission similar to that of the $\rm 150\, MHz$ counterpart. The principal causes of the ro-vibrational transitions of $\rm H_2$ are the fluorescent excitation ("UV pumping") and the processes associated with thermal causes (collisions). In the work of \citet{Moorwood}, they conclude that in a starburst galaxy, as is the case of Arp\,299, the most plausible scenario is the one that favors the collisions, for example, the shock of an outflow with the interstellar medium (ISM). On the other hand, the existence of [FeII] emission could be associated with both compact objects or extended emission. The SNe are responsible for the compact emission, coming from the ionized medium after the shocks produced in the explosions propagate through the ISM, while in the form of extended emission is directly related with nuclear outflows \citep{Greenhouse,Alonso-Herrero2003}. The image of FeII presents a combination of biconical structure and extended emission following a spiral pattern. As pointed out in the previous subsection, this can be explained as being due to the Southern part being covered by the  ongoing starburst in the galactic disk, which distorts or generates emission not directly related with the outflow but with the spiral arms. Therefore, here we focus on the Northern part to avoid any potential FeII emission from SN shocks inherent to the clumpy disk. The bulk of the emission originates in the nucleus, but there is a very faint structure to the North that follows the same shape as the putative outflow and that is brighter at the edges, thus resembling filaments.This enhanced emission may be the result of FeII emitter stacking, the trigger of a process that favors the FeII or both scenarios combined. As Arp\,299-A is an ongoing factory of SNe \citep{Perez-Torres09}, one might ask if the FeII extended emission is the sum of all the generated compact FeII emission plus the winds associated in the region. The presence of this emission up to $\rm 2.5\, kpc$ away from the nucleus, where the bulk of SNe live, cannot be explained by an emission generated by shocked material in the nucleus and transported 
northwards, emitting Fe~II on its way towards the regions of enhanced emission. Even though the outflow is most likely generated by the SN factory present in the nucleus, the Fe~II emission observed in the Northern component is most likely coming from shocks produced by the outflow itself. 

The filament in the FeII image observed in the NW region matches the edge of the emission in both $\rm H_2$ and at $\rm 150\, MHz$. The co-existence of $\rm H_2$ and FeII in the same region indicates that the process behind these lines is an outflow that produces shocks with the ISM in its wake \citep{Moorwood}.

\begin{figure}
\includegraphics[width=1.\linewidth]{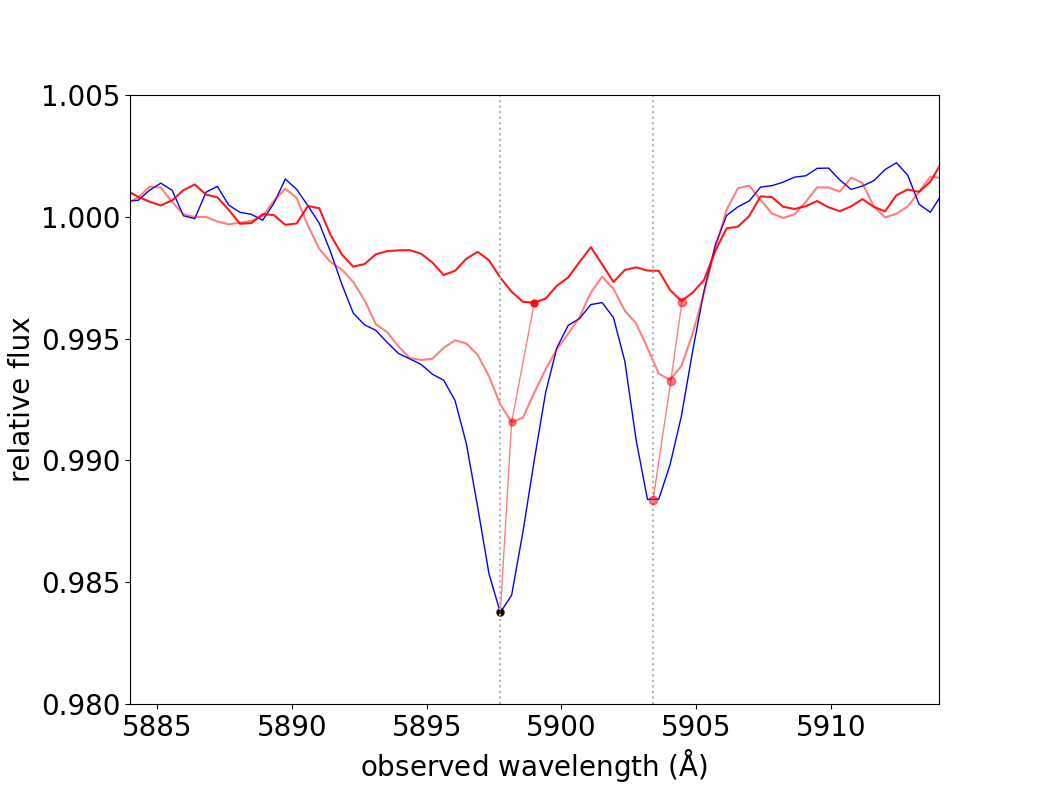}
\caption{NaI D lines in the nucleus (blue) and in two adjacent fibers corresponding to outer regions (red). The outer profiles are blueshifted indicating an outflow. Dots indicate the local minimum of the line. Dashed vertical lines indicate the reference wavelength positions of the absorption lines in the nucleus. The spectra is flux normalized in the continuum window 5805 - 5930 \AA.}
\label{fig:NaI}
\end{figure}

\textit{Na\,I absorption in the nucleus:} Because of its low ionization potential, the resonant NaI D lines are a good tracer of the neutral ISM. NaI absorption was used in the pioneering work of \citet{Phillips} as a probe of a superwind in NGC 1808, and in a systematic study of 32 FIR bright starburst galaxies by \citet{Heckman}. Moreover, these double sodium absorption lines have been used in the study of outflows in ULIRGS \citep{Rupkea,Rupkeb}. Figure \ref{fig:NaI} shows the spectra around the NaI D lines $\lambda\lambda$5889.95, 5895.92 extracted from the nucleus (blue line) and from an adjacent region (in red). The absorption is redshifted in the outer regions, indicating a large amount of outflowing dense neutral medium.  

The spectra are flux normalized in the continuum window 5805-5930 \AA.

\subsection{The nature of the outflow in Arp\,299-A}

There is little doubt that the LOFAR 150 MHz morphology of the extended emission in the central regions of Arp\,299-A, in addition to the morphology of the Fe II and H$_2$ near-IR  observations and the Na\,I D spectra, signals the existence of an outflow in the nucleus of Arp\,299-A. Arp\,299-A is known to host a low-luminosity active galactic nuclei (LLAGN, \citealt{Perez-Torres10,Alonso-Herrero2013}) defined as an AGN with $\rm L(H\alpha) \leq 10^{40} erg\, s^{-1}$ \citep{Ho}, although its precise nature  (Seyfert or LINER) is still unknown. Further, it also presents a powerful starburst \citep{Perez-Torres09,Ulvestad,Bondi} in its nuclear region. Thus, a natural question arises regarding the mechanism that is powering the outflow. Is it the LLAGN in Arp299-A, or is it a compact starbust?

AGNs can drive outflows in the central regions of galaxies, as their accretion power can be very high (e.g., \citealt{Veilleux}):

\begin{equation}
  \dot{E}_{\rm acc} \simeq 10^{11}\, \epsilon_{-2}\, \dot{M}_{\rm acc}\, L_\odot,  
\end{equation}

where $\rm \dot{M}_{\rm acc}$ is the mass accretion rate in $\rm M_{\odot} yr^{-1}$ and 
$\rm \epsilon_{-2} = \epsilon/0.01$ is the energy conversion efficiency in rest mass units. 
The mass accretion rate ranges from $\la$ 0.001
M$_\odot$~yr$^{-1}$ for low-luminosity AGN, $\sim$ 1
M$_\odot$~yr$^{-1}$ for Seyfert galaxies, and $\sim$ 100
M$_\odot$~yr$^{-1}$ for quasars and powerful radio galaxies \citep{Veilleux}.

EVN observations of Arp\,299-A showed that Arp\,299-A hosts an LLAGN at its center \citep{Perez-Torres10}, with $\rm L_{bol} \approx 6.4 \pm  0.6 \times 10^{43} erg s^{-1}$ \citep{Alonso-Herrero2013}, surrounded by a large core-collapse supernova factory \citep{Perez-Torres09}. 
The accretion power of the LLAGN in Arp299-A is unlikely to be larger than $\rm \dot{E}_{\rm acc} \simeq 3.8\times 10^{41} \epsilon_{-2}$ erg~$s^{-1}$, but still might produce outflow phenomena as is the case in, for example, NGC\, 1433 \citep{combes13}, or NGC\, 1068 \citep{Garcia-Burillo}.

We should note that one major problem with the scenario of a LLAGN-driven outflow for Arp299-A is the orientation of the jet.  We know from our previous EVN observations \citep{Perez-Torres10} that the radio jet  emanating from the AGN in Arp\,299-A is oriented along the East-West direction, while the outflow 
unveiled by LOFAR is in the North-South direction, almost perpendicular (in the projected plane) to the AGN jet.

The main alternative scenario is that of a starburst driving the outflow, given the existence of intense supernova activity in the central $\rm \sim 150$ pc \citep{Perez-Torres09, Bondi}. In the early stages of a starburst, that is, before core collapse supernovae (CCSNe) start exploding, the main input of mechanical energy into the ISM is through stellar winds, in particular those of Wolf-Rayet stars \citep{Leitherer}. However, soon after CCSNe come into play, their mechanical input dominates for starburst ages $\rm \geq 6$ Myr (see, e.g., Figs. 108 and 114 of \citet{Leitherer}. Given that we have direct observational evidence of ongoing CCSNe explosions in the central $\sim$150 pc of Arp\,299-A from our own VLBI observations \citep{Perez-Torres09,Bondi} and an estimation of 10 - 15 Myr for the starburst age \citep{Alonso-Herrero}, we consider here only the mechanical input from CCSNe, for the sake of simplicity. 
\citet{Bondi} found a lower limit to the CCSNe rate in Arp\,299-A of $\rm r_{SN} \geqslant 0.8 SN yr^{-1}$. 
Assuming standard values for the energy of CCSN explosions, $\rm E_{SN} = 10^{51}\,E_{51}$ erg (where $\rm E_{51}$ is the energy in units of $\rm 10^{51}$ erg),
and for the conversion efficiency of the energy of the explosion into mechanical energy, $\rm \eta = 0.1\,\eta_{-1}$ \citep{Thornton}, we obtain 
$\rm \dot{E}_{sb} \simeq 3.2\times 10^{42} r_{SN,1}\, \eta_{-1}$ erg~$s^{-1}$, where 
$\rm r_{SN,1} = r_{SN}/yr^{-1}$. The available mechanical energy due to SN activity is almost ten times larger than that due to AGN activity in Arp\,299A, and hence a starburst can account for the observed outflow much more easily than the LLAGN. 

We can also estimate the maximum outflowing mass-loss rate of a starburst-driven outflow, as follows: 

\begin{equation}
 \dot{M}_{\rm sb} \simeq 1000 \, \eta_{-1}\, E_{51}\, r_{\rm SN,1}\, v^{-2}_2\, M_{\odot} \rm yr^{-1},
 \label{eq:Mdot_sb}
\end{equation}
 
where $\rm v_2 = v/100\, {\rm km s^{-1}}$ is the outflow speed, in units of $\rm 100\, km s^{-1}$. 
Starburst-driven outflows show wind speeds of about $\rm 400 km s^{-1}$ \citep{Veilleux}, so the maximum outflowing mass rate supported by the starburst is $\rm \dot{M}_{sb} \simeq 63 M_{\odot}  yr^{-1}$. We can independently obtain another estimate for the starburst mass outflow rate, using the approximation in \citet{Veilleux}:

\begin{equation}
\dot{M}_{\rm sb}  = 0.26~(\rm SFR / M_\odot~{\rm yr}^{-1})~~M_\odot\ {\rm yr}^{-1} 
,\end{equation}

\noindent where SFR is the star formation rate ($\rm M_{\odot} yr^{-1}$), solar-metallicity is assumed, 
and the starburst age is beyond $\sim$ 40 Myr, implying  
 that the  mass-loss rate and mechanical luminosity are constant. Assuming, for simplicity, a constant SFR, the CCSN rate for a Salpeter IMF is approximately $\rm r_{sn} \approx 0.019\, SFR\, \rm yr^{-1}$ (e.g., Eq. 9 in \citealt{Perez-Torres09b}), and the above expression becomes 
$\rm \dot{M}_{sb} \simeq 13.6\, r_{SN,1}\, M_{\odot}yr^{-1}$, or about 11 $\rm M_{\odot} \,yr^{-1}$ for the case of Arp\,299A.

We can now rewrite Eq. \ref{eq:Mdot_sb} to obtain the outflow speed:

\begin{equation}
\rm v \simeq 3200 \, \eta_{-1}\, E_{51}\, r_{\rm SN,1},\dot{M}_{\rm sb}^{-1} \rm km\,s^{-1}
,\end{equation}

where $\rm \dot{M}_{sb}$ is in units of $\rm M_{\odot}yr^{-1}$. 
We thus obtain the range of mass outflow rates $\rm \simeq 11 - 63 M_{\odot}yr^{-1}$. Substituting those values into the previous equation, and using $\rm r_{SN} \geqslant 0.8 yr^{-1}$ from our VLBI observations, we obtain outflow speeds in the range $\rm v \simeq 370 - 890 km s^{-1}$, which are in broad agreement with observed wind speeds in 
starburst-driven outflows (e.g., \citealt{Veilleux}). 

The range of velocities estimated above, in addition to the physical size of the outflow and the position of the nuclear starburst can help to get an idea of the age of the outflow.  However, this length is affected by the projection on the sky of the real outflow, so, primarily, we must deproject it (HyperLeda \footnote{HyperLeda: http://leda.univ-lyon1.fr} angle value is $i \approx 52^\circ$). Therefore, the total size of the outflow would be $\rm \approx 5\, kpc$. Since the compact starburst in Arp299-A
surrounds the A nucleus \citep{Perez-Torres09}, the outflow radius is then $\rm 2.5\, kpc$.
For the range of velocities obtained above ($\rm 370-890\, km\,s^{-1}$), the kinematic age of the outflow is ($3-7$) Myr.

\section{Summary and conclusions}\label{sec:summary}

We have presented for the first time highly sensitive, sub-arcsecond ($\rm \sim0.4$ arcsec) LOFAR observations at $\rm 150\, MHz$ of the merging galaxy system Arp\,299, focusing on the Arp\,299-A nucleus.

Our main finding is the presence of a two-sided, wide filamentary structure emanating from the A-nucleus, which we interpret as an outflow that extends up to at least $\rm \simeq 3$ kpc from the A-nucleus along the N-S direction.
Published HST/NICMOS [FeII]$\rm \, 1.64\mu m$ and H$_2$ at 2.12$\rm \,\mu m$ images of Arp\,299-A \citep{Alonso-Herrero} show very similar features to those obtained by us with LOFAR at 150 MHz, providing further morphological support for the putative outflow. Finally, previously unpublished Na\,I D spectra show a Doppler shift in this absorbed lines between the A nucleus and other two surrounding regions, also reinforcing the existence of an outflow in Arp\,299-A.  
 
We  discuss  two main scenarios to power the outflow, namely that of a starburst, or that of an AGN. The mechanical energy available from the starburst in the central regions of Arp\,299-A is very large, $\rm \dot{E}_{sb} \simeq 3.2\times 10^{42} r_{SN,1}\, \eta_{-1}$ erg~$s^{-1}$ due to the large core-collapse supernova rate \citep{Bondi}. On the other hand, the mechanical energy due to the accretion power of the LLAGN sitting at the center of Arp\,299-A is almost ten times smaller.  We thus conclude that the outflow in Arp\,299-A is driven by the powerful starburst in its central regions.
We estimate that the starburst wind can support a mass-outflow rate in the
range (11 - 63) $\rm M_{\odot} yr^{-1}$ at speeds of up to (370 - 890) $\rm km ^{-1}$, and that it is relatively young, with an estimated kinematic age of $\rm (3 - 7)$ Myr.

Our results indicate that sub-arcsecond imaging with LOFAR may be an excellent new tool for unveiling outflows in the central regions of local LIRGs.

\bibliography{bibtex/astroph}

\clearpage

\appendix
\counterwithin{figure}{section} 
 
\section{LOFAR at 150\, MHz}
Observations were taken in an 8-bit mode with the high band antenna (HBA) part of LOFAR in dual\textunderscore inner configuration. This means that each of the two "ears" of the 24 core stations were correlated as separated stations. This provides many more short baselines between core stations, which enable future legacy studies of extended emission on much larger scales than we study in this paper. A total of 71 LOFAR stations participated: 48 core stations (CS), 14 remote stations (RS) and 9 international stations (IS). Unfortunately the data from the Swedish station SE607HBA could not be calibrated and were excluded from our analysis \footnote{Although the delays and phases calibrate well, the visibility amplitudes in the upper third quartile of the data channels from SE607HBA cannot be calibrated. Due to the beam configuration in LC5\textunderscore 20, calibration transfer is not possible for this station. This issue was first reported for cycle 2 data by \citet{varenius2016} and seen in cycle 7 data from March 2017.}.
\subsection{Pre-averaging and flagging}
The LTA stores data as one compressed  measurement set (MS) per subband (162) per source (4) per scan (24; 30 min, or 3 min for 3C~295), that is, about 2.8~TB of data spread over about 15500 files. After downloading all files from the LTA we used LOFAR Default Processing Pipeline (DPPP) to apply the LOFAR beam model, combined all files in frequency for each scan and source, and averaged further to 8~sec in time and 195~kHz per channel to reduce processing time. This limits the smearing radius to about $1'$ for a 1000~km baseline. We ran the standard LOFAR software AOFlagger v2.8 \citep{offringa-2012-morph-rfi-algorithm} with default parameters to remove radio frequency interference (RFI) signals from the data. We then used DPPP to derive phase-corrections for all CS using a point source model of 3C~295, using the scans on 3C~295. The corrections were extended in time to allow application of the derived corrections to the interleaved scans of Arp\,299 and the two calibrators. After applying the phase corrections, thereby aligning the phases of the CS, we used DPPP to combine stations into super-stations to increase processing time and baseline sensitivity. CS 2,3,4,5,6 and 7 were joined together into a "super terp" station (ST). For the remaining CS we joined the "ears" together for each station. Hence we produced an averaged and summed data set with 41 stations: 14 RS, 8 IS, 18 CS and one phased ST. Next, we used CASA 5.1.1 task \verb!concat! to concatenate the scans to one single MS per source. These four MSs were converted from linear to circular polarization using the Table Query Language (TAQL) utility \verb!mscal.stokes!. Finally, the task \verb!exportuvfits! in CASA was used to export the data to UVFITS-format for processing in the NRAO Astronomical Image Processing System (AIPS release 31DEC16; see e.g., \citealt{AIPS}). We used the ParselTongue Python interface \citep{parseltongue} version 2.3 to script AIPS tasks.

\subsection{Calibration and imaging}
The AIPS calibration strategy was very similar to the one used for M~82 and Arp\,220 by \citet{varenius2015,varenius2016}, and therefore only the key points are summarized here. Residual delays, rates and amplitude errors were calibrated using AIPS towards J1128+5925, assuming this source to be 270~mJy with flat spectrum throughout the 32~MHz observing band. The data were split into four 8-MHz spectral windows to enable a nondispersive (linear) correction for phase slopes versus frequency using the task FRING in AIPS with a minimum baseline length of 60~k$\rm \lambda$. We used a solution interval in FRING of 1~min for IS and 2~min for RS. After FRING the data were averaged to 24~sec in time. Bandpass corrections were derived towards J1128+5925 once every hour and the data were then averaged to 780~kHz per channel for calibrators and 390~kHz per channel for Arp\,299, which at this point implies a 10\% smearing radius of about $\rm 20''$ and $\rm 35''$ , respectively, for 1000~km baselines. An iterative self-calibration loop was used to calibrate the relative amplitude scale using J1128+5925 to account for possible source structure. The calibration was transferred to 1127+5841 where the phases were refined assuming a point source model, and all cumulative corrections were applied to Arp\,299. The corrections derived towards J1128+5925 were also transferred to 3C~295 to check the absolute flux density scale. Finally, Arp\,299 was imaged using the CLEAN deconvolution algorithm as implemented in the AIPS task \verb!IMAGR!, including baselines longer than 5~k$\rm \lambda$ to sample also the extended emission.

   \begin{table}
      \caption[]{Coordinates of calibrators and target referred to in Sect. \ref{sec:reduction}.}
         \label{tab:targetlist}
         \begin{tabular}{ l l l}
            Source      &  R. A. [J2000] & Dec. [J2000]\\
            \hline
            J1127+5841 \tablefootmark{a} & $11^{\rm h}27^{\rm m}34^{\rm s}.4620$ & $58^\circ41'41\farcs821$\\
            J1128+5925 \tablefootmark{b} & $11^{\rm h}28^{\rm m}13^{\rm s}.3407$& $59^\circ25'14\farcs799$\\  
            Arp\,299  \tablefootmark{c} & $11^{\rm h}28^{\rm m}32^{\rm s}.25$ & $58^\circ33'42\farcs0$ \\            
            3C~286 \tablefootmark     &$13^{\rm h}31^{\rm m}08^{\rm s}.2881$ & $30^\circ30'32\farcs959$\\
            3C~295 \tablefootmark{d}     & $14^{\rm h}11^{\rm m}20^{\rm s}.50$ & $52^\circ12'10\farcs0$\\
            \noalign{\smallskip}
            \hline
         \end{tabular}
         \tablefoot{
             \tablefoottext{a}{From the VLA L-band project AA0216.}
             \tablefoottext{b}{From the \emph{rfc\_2015d} catalogue available
             via http://astrogeo.org/calib/search/html.}
             \tablefoottext{c}{LOFAR correlation position, between Arp\,299A and Arp\,299B.}
             \tablefoottext{d}{From NED, http://ned.ipac.caltech.edu.}

}
\end{table}

\begin{figure}
\includegraphics[width=1.\linewidth]{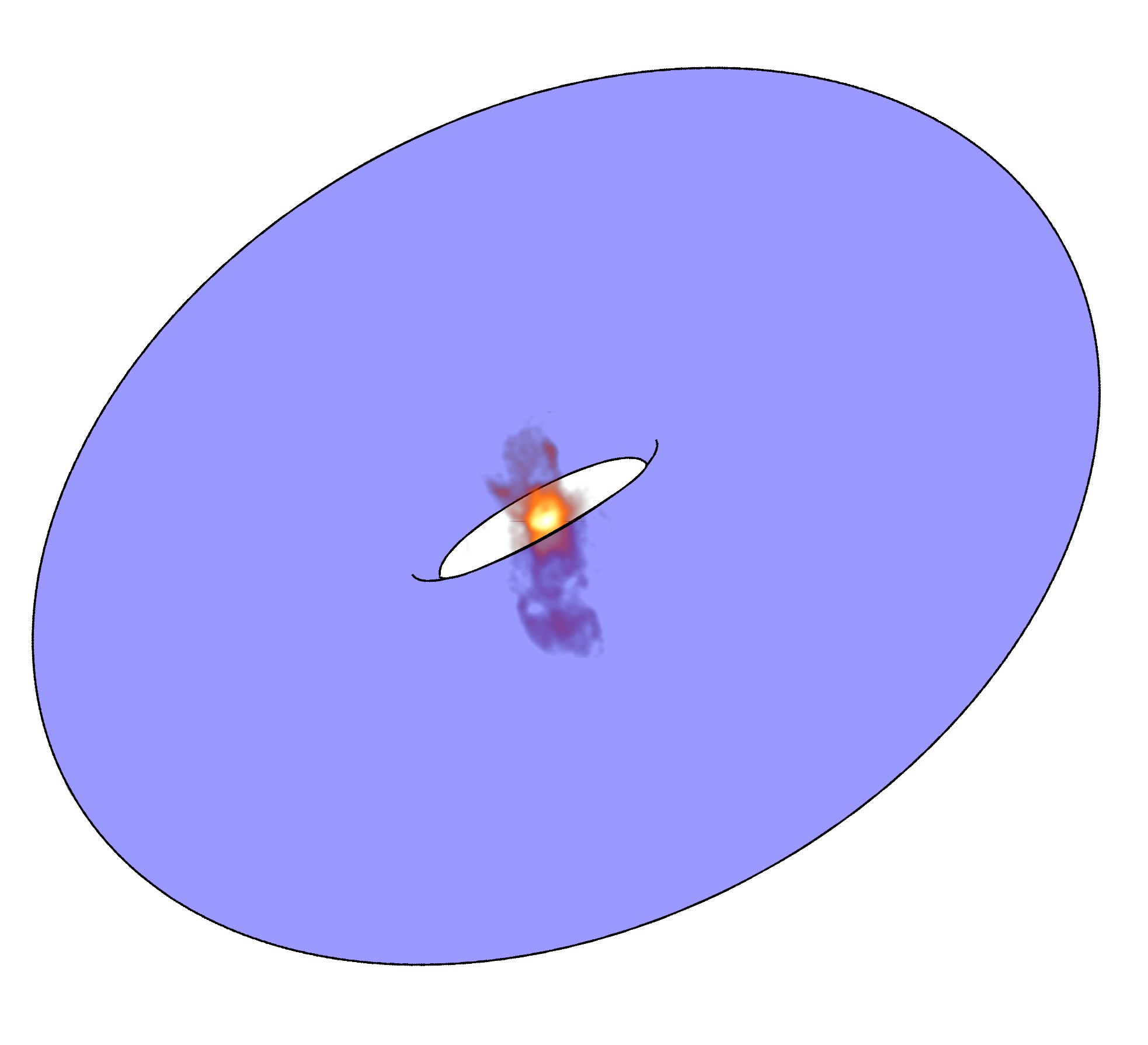}
\caption{Sketch of the outflow in Arp\,299 A (IC694), where the emulated disk covers around 60 arcseconds in the major axis and the inclination of the disk is $\approx 30^{\circ}$. The clouds are located in the galactic disk (dark blue torus), thus obscuring the southern part of the outflow.
The sketch tries also to represent the real inclination of the galactic disk.
}
\label{fig:toy_model}
\end{figure}

\begin{figure}[htbp]
\centering
\subfigure[]{\includegraphics[width=43mm]{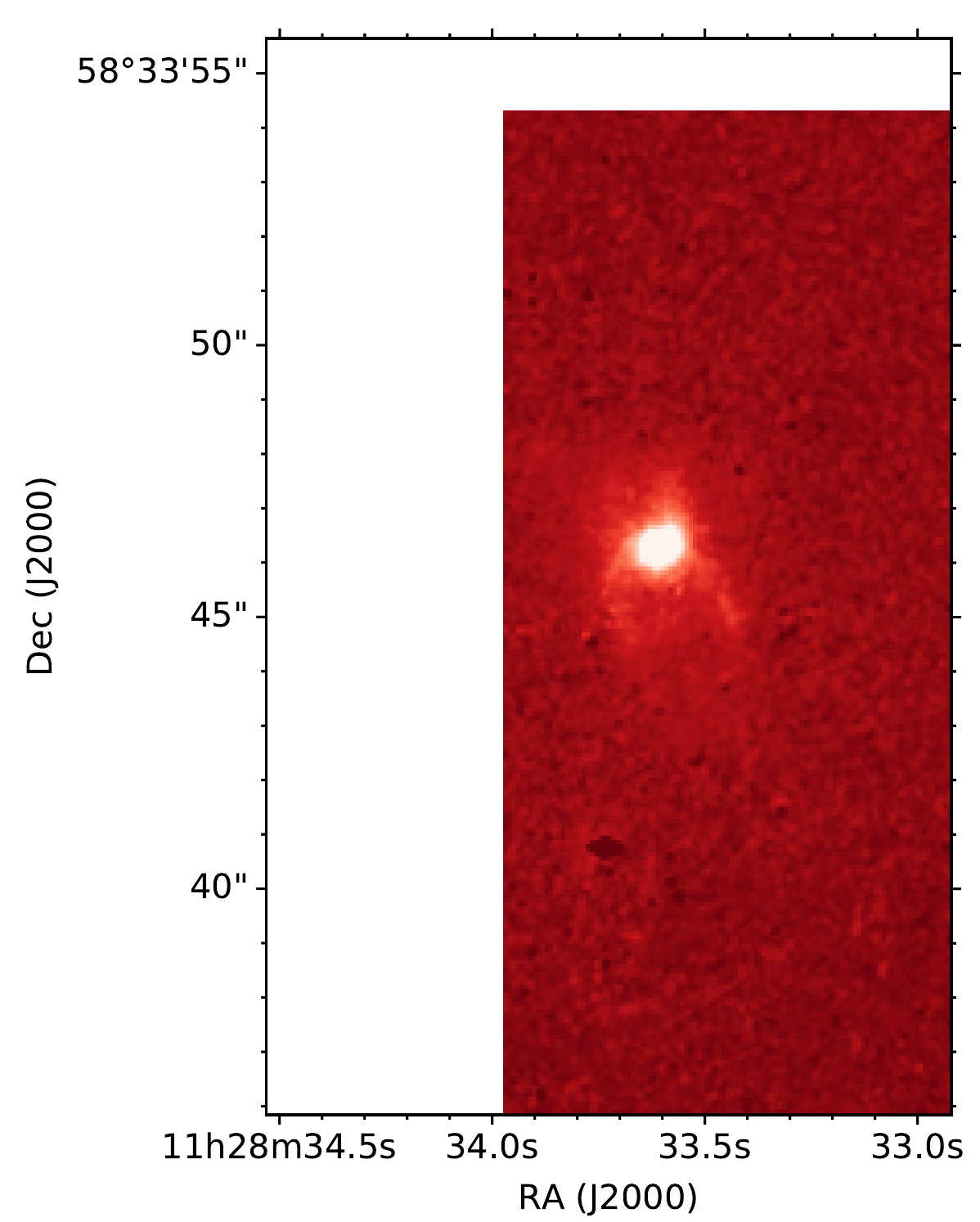}}
\subfigure[]{\includegraphics[width=43mm]{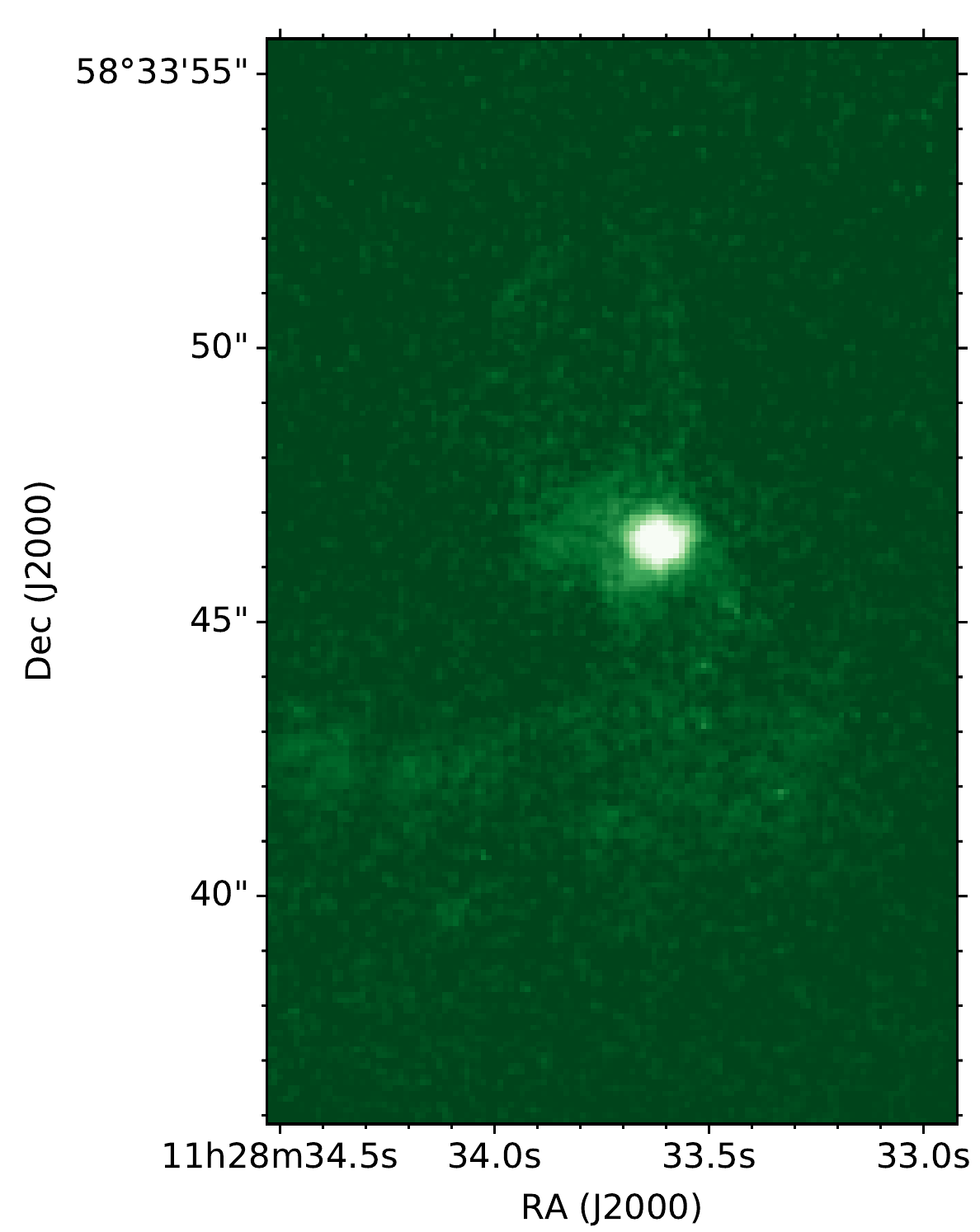}}
\subfigure[]{\includegraphics[width=80mm]{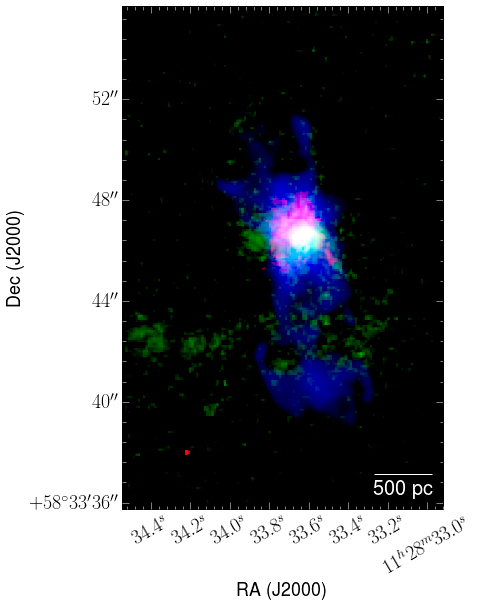}}
\caption{Images of (a) $\rm H_2$ and (b) FeII from \citet{Alonso-Herrero}. H2 is represented in linear scale, while FeII is in logarithmic scale. The H2 image  is blanked in the left and upper regions since those regions were not covered by NICMOS observations. The composite RGB image (c) of the same region of Arp\,299-A $\rm H_2$ and FeII , and the LOFAR image at $\rm 150\, MHz$, in red, green and blue, respectively.} \label{fig:RGB_FeIIH2}
\end{figure}

\label{appendix}


\end{document}